\documentclass[12pt]{article}
\usepackage{epsfig}

\psfigdriver{dvips}
\renewcommand{\baselinestretch}{1.0}
\def\bbox#1{%
\relax\ifmmode
\mathchoice
{{\hbox{\boldmath$\displaystyle#1$}}}%
{{\hbox{\boldmath$\textstyle#1$}}}%
{{\hbox{\boldmath$\scriptstyle#1$}}}%
{{\hbox{\boldmath$\scriptscriptstyle#1$}}}%
\else
\mbox{#1}%
\fi
}

\begin{document}

{\Large\bf\centerline{Systematic Renormalization in Hamiltonian Light-Front}}
{\Large\bf\centerline{Field Theory:  The Massive Generalization}}
\vspace{.2in}

\centerline{Roger D. Kylin\footnote{E-mail:  kylin@mps.ohio-state.edu}, Brent H. Allen\footnote{E-mail:  
allen@mps.ohio-state.edu}, and Robert J. 
Perry\footnote{E-mail:  perry@mps.ohio-state.edu}}

\vspace{.1in}
\centerline{\it Department of Physics, The Ohio State University, Columbus, Ohio 43210}
\vspace{.5cm}

\centerline{\today}

\begin{quote}
\renewcommand{\baselinestretch}{1.0} 

\abstract{Hamiltonian light-front field theory can be used to solve for hadron states in QCD.  To this end, a method has been 
developed for systematic renormalization of Hamiltonian light-front field theories, with the hope of applying 
the method to QCD.  It assumed massless particles, so its immediate application to QCD is limited to gluon 
states or states where quark masses can be neglected.  This paper builds on the previous work by including particle 
masses non-perturbatively, which is necessary for a full treatment of QCD.  We show that several subtle new 
issues are encountered when including masses non-perturbatively.  The method with masses is algebraically and conceptually 
more difficult; however, we focus on how the methods differ.  We demonstrate the method using massive $\phi^{3}$ theory 
in 5+1 dimensions, which has important similarities to QCD.}
\end{quote}

\vskip .25in
\renewcommand{\baselinestretch}{1.0}
\tiny
\small
\section{Introduction}
The use of a Hamiltonian light-front formalism may simplify the solution of quantum chromodynamics (QCD) by allowing 
us to make a convergent expansion of hadron states in free-particle Fock-space sectors.  The Fock-space expansion 
will rapidly converge if the Hamiltonian satisfies certain conditions \cite{brent}. The Hamiltonian can then be used to 
solve for approximate hadron states.  

Inspired by the work of Dyson 
\cite{dyson}, Wilson \cite{wilson}, G{\l}azek and Wilson \cite{glazek and wilson}, and Wegner \cite{wegner}, significant 
work has been done to perturbatively derive light-front Hamiltonians that satisfy these conditions in the 
full Fock-space, neglecting zero modes 
[6-13].  
The method developed by Allen and Perry \cite{brent} includes the scale dependence of the coupling and can be used to 
systematically renormalize light-front Hamiltonians, fixing all non-canonical operators, in principle to all orders.
 
In this method the theory is regulated by placing on the Hamiltonian a smooth cutoff on change in free mass.  
The cutoff violates several physical principles, preventing 
renormalization exclusively through the redefinition of masses and canonical couplings. Renormalization 
must be completed by requiring the Hamiltonian to produce cutoff-independent physical quantities, and by requiring it 
to obey the physical principles of the theory that are not violated by the cutoff.  These requirements 
completely fix the Hamiltonian so that it will give results consistent with all the physical principles of the theory, even those 
violated by the cutoff.  The most powerful characteristic of this approach is that it systematically ``repairs'' the 
theory and retains only the fundamental parameters of the theory.

We generalize this method of renormalization to include particle masses and demonstrate it using massive $\phi^{3}$ 
theory in 5+1 
dimensions.  This theory is asymptotically free and its diagram structure is similar to QCD, which make it a good 
perturbative development ground.  It is straightforward to extend the method for massless theories developed in Ref. \cite{brent} 
to calculate QCD quantities for which particle masses are unimportant, such as the low-lying glueball spectrum 
\cite{brent2}.  In this paper, we show how to incorporate particle masses non-perturbatively as a necessary step toward 
a treatment of full QCD.

 \section{Review of General Formalism}
 \label{sec:formal}
In this section we introduce some of the notation developed in Ref. \cite{brent} and outline the method.
Formalism that is necessary for a detailed understanding of this method but that we do not repeat in this paper 
can be found in this earlier work.  This includes the use of a unitary transformation to determine how the 
Hamiltonian changes with the cutoff \cite{wegner}, the use of physical principles to restrict the form of the Hamiltonian, and the 
details of how to compute matrix elements of the Hamiltonian.
 
We want to find the regulated invariant-mass operator, ${\cal M}^{2}(g_{_{\Lambda}},m,\Lambda)$, which is trivially related to the 
Hamiltonian.  It can be split into a free part (which contains implicit mass dependence) and an interacting part:
\begin{equation}
\label{separation}
{\cal M}^{2}(g_{_{\Lambda}},m,\Lambda)={\cal M}^{2}_{\mathrm{free}}+{\cal M}_{\mathrm{int}}^{2}(g_{_{\Lambda}},m,\Lambda).
\end{equation}
Since the method treats ${\cal M}_{\mathrm{int}}^{2}(g_{_{\Lambda}},m,\Lambda)$ perturbatively, we put the particle-mass term in ${\cal 
M}_{\mathrm{free}}^{2}$, to treat it non-perturbatively; however, ${\cal M}_{\mathrm{int}}^{2}(g_{_{\Lambda}},m,\Lambda)$ will still have mass dependence.
The matrix elements of ${\cal M}^{2}(g_{_{\Lambda}},m,\Lambda)$ are written
\begin{eqnarray}
\langle F \vert {\cal M}^{2}(g_{_{\Lambda}},m,\Lambda)\vert I\rangle&=&\langle F\vert {\cal M}_{\mathrm{free}}^{2}\vert I\rangle+
\langle F\vert {\cal M}_{\mathrm{int}}^{2}(g_{_{\Lambda}},m,\Lambda)\vert I\rangle\nonumber\\
&=&M^{2}_{F}\langle F\vert I\rangle +e^{-\frac{\Delta^{2}_{FI}}{\Lambda^{4}}}\langle F\vert V(g_{_{\Lambda}},m,\Lambda)\vert 
I\rangle,
\end{eqnarray}
where $\vert F\rangle$ and $\vert I\rangle$ are eigenstates of the free invariant-mass operator with eigenvalues 
$M_{F}^{2}$ and $M_{I}^{2}$, and $\Delta_{FI}$ is the difference of these eigenvalues.  $V(g_{_{\Lambda}},m,\Lambda)$ is the 
interacting part of the invariant-mass operator with the Gaussian cutoff factor removed 
and is called the ``reduced interaction.''  With light masses $\Delta_{FI}$ will be small in a limited part of 
phase space.  This means there will be non-perturbative effects that must be dealt with through the diagonalization 
of the mass operator rather than its perturbative renormalization.

We expand $V(g_{_{\Lambda}},m,\Lambda)$ in powers of the running coupling,
 $g_{_{\Lambda}}$:
\begin{equation}
\label{gseries}
V\hspace{-.07cm}(g_{_{\Lambda}},m,\Lambda)=\sum_{r=1}^{\infty}g_{_{\Lambda}}^{r}V^{(r)}\hspace{-.07cm}(m,\Lambda),
\end{equation}
where $V^{(1)}$ is the canonical interaction and the $V^{(r\ge 2)}\hspace{-.07cm}(m,\Lambda)$'s are non-canonical 
interactions.  These non-canonical operators can be thought of as counterterms in a traditional approach.  
Note that $g_{_{\Lambda}}$ implicitly depends on $m$.  We use a unitary transformation to 
relate ${\cal M}^{2}(g_{_{\Lambda}},m,\Lambda)$ to ${\cal M}^{2}(g_{_{\Lambda^{\prime}}},m,\Lambda^{\prime})$, 
where $\Lambda^{\prime} > \Lambda$.  This yields the relation
\begin{equation}
\label{diff}
V^{(r)}\hspace{-.07cm}(m,\Lambda)-V^{(r)}\hspace{-.12cm}\left(m,\Lambda^{\prime}\right)=
{\delta V^{(r)}\hspace{-.07cm}(m,\Lambda,\Lambda^{\prime})-\sum_{s=2}^{r-1}B_{r-s,s}V^{(r-s)}\hspace{-.07cm}(m,\Lambda)},
\end{equation} 
where $\delta V^{(r)}\hspace{-.07cm}(m,\Lambda,\Lambda^{\prime})$ is the ${\cal O}(g^{r}_{_{\Lambda^{\prime}}})$ 
change in the reduced interaction and 
the $B_{r-s,s}$'s are functions of $m$, $\Lambda$, and $\Lambda^{\prime}$ that contain information on the scale dependence 
of the coupling.  Since the scale dependence of the reduced interaction comes from $g_{_{\Lambda}}$ and the 
$V^{(r)}\hspace{-.07cm}(m,\Lambda)$'s [See Eq. (\ref{gseries})], Eq. (\ref{diff}) simply states that if we subtract 
from $\delta V^{(r)}\hspace{-.07cm}(m,\Lambda,\Lambda^{\prime})$ the contribution due to the scale dependence of the coupling, then we are left with the 
contribution due to the scale dependence of the $V^{(r)}\hspace{-.07cm}(m,\Lambda)$'s.

If there is a part of $V^{(r)}\hspace{-.07cm}(m,\Lambda)$ that is independent of the cutoff, it will cancel on the 
left-hand-side of Eq. (\ref{diff}).  For this reason, we split $V^{(r)}\hspace{-.07cm}(m,\Lambda)$ into a part that 
depends on the cutoff, $V^{(r)}_{\mathrm{CD}}\hspace{-.07cm}(m,\Lambda)$, and a part that is independent of the cutoff, 
$V^{(r)}_{\mathrm{CI}}\hspace{-.07cm}(m)$:
\begin{equation}
V^{(r)}\hspace{-.07cm}(m,\Lambda)=
V^{(r)}_{\mathrm{CD}}\hspace{-.07cm}(m,\Lambda)+V^{(r)}_{\mathrm{CI}}\hspace{-.07cm}(m).
\end{equation}
This division can be made with no ambiguity because we are assuming approximate transverse locality.
Solving for both $V_{\mathrm{CD}}^{(r)}\hspace{-.07cm}(m,\Lambda)$ and $V_{\mathrm{CI}}^{(r)}\hspace{-.07cm}(m)$ is 
necessary to find the invariant-mass operator.

\section{Addition of Particle Masses}
\label{sec:difference}
In our renormalized scalar theory $m$ is the physical particle mass to all orders in perturbation theory.  In a 
confining theory $m$ would be considered to be the particle mass in the 
zero-coupling limit.  Since the mass is being treated non-perturbatively, it must be included in the free part of 
${\cal M}^{2}(g_{_{\Lambda}},m,\Lambda)$ in Eq. (\ref{separation}).  This leads to an altered unitary transformation and 
fundamental changes in the renormalization procedure.  

The changes in the procedure are discussed in the next three subsections.  The redefinition 
of the coupling (Sec. \ref{coupling}) is straightforward.  In Sections \ref{dependent} and \ref{independent}, 
we present the expressions for the matrix elements of $V_{\mathrm{CD}}^{(r)}\hspace{-.07cm}(m,\Lambda)$ and 
$V_{\mathrm{CI}}^{(r)}\hspace{-.07cm}(m)$, respectively.  We also qualitatively discuss the additional steps that are required to 
interpret and use them in a massive theory.

\subsection{Coupling}
\label{coupling}
The canonical definition of the coupling is
\begin{equation}
g = \left[ 64 \pi^{5} p_1^+ \delta^{(5)}(p_1 - p_2 - p_3) \right]^{-1} \left< \phi_2 \phi_3 \right| 
{\cal M}^2_{\mathrm{can}}
\left| \phi_1 \right>|_{p_{2}=p_{3}} .
\end{equation}
In the massive theory, we choose
\begin{eqnarray}
g_{_{\Lambda}}&=&\left[64\pi^{5}p_{1}^{+}\delta^{(5)}\left(p_{1}-p_{2}-p_{3}
\right)\right]^{-1}
\mathrm{exp}\left(9\frac{m^{4}}{\Lambda^{4}}\right)\langle 
\phi_{2}\phi_{3}\vert{\cal{M}}^{2}(g_{_{\Lambda}},m,\Lambda)
\vert\phi_{1}\rangle|_{p_{2}=p_{3}}\nonumber\\
&=&\left[64\pi^{5}p_{1}^{+}\delta^{(5)}\left(p_{1}-p_{2}-p_{3}
\right)\right]^{-1}
\langle \phi_{2}\phi_{3}\vert 
V(g_{_{\Lambda}},m,\Lambda)\vert\phi_{1}\rangle|_{p_{2}=p_{3}},
\end{eqnarray}
which differs from the definition in the massless theory by the factor 
$\mathrm{exp}\left(9\frac{m^{4}}{\Lambda^{4}}\right)$.  This choice of coupling cancels a cutoff factor due to the 
presence of the mass and allows us to closely follow the formalism developed in the massless theory.  In 
particular, the expressions for the matrix elements of $V_{\mathrm{CD}}^{(r)}\hspace{-.07cm}(m,\Lambda)$ and 
$V_{\mathrm{CI}}^{(r)}\hspace{-.07cm}(m)$ presented below have the same form as those derived in Ref. \cite{brent}.

\subsection{Cutoff-Dependent Contributions to $\bbox{V^{(r)}\hspace{-.07cm}(m,\Lambda)}$}
\label{dependent}

Momentum conservation implies that any matrix element of $V^{(r)}\hspace{-.07cm}(m,\Lambda)$ contains a sum of terms, 
each with a unique product of momentum-conserving delta functions.  Assuming that approximate transverse locality is 
maintained, the coefficient of each product of delta functions can be written as an 
expansion in powers of transverse momenta.  In massive $\phi^{3}$ theory, we can also make a generalized expansion in 
powers and logarithms of $m$.  The scale dependence of any term in this expansion has the form
\begin{equation}
\label{fcdi}
\Lambda^{6-2N_{\mathrm{int}}}\left(\frac{m}{\Lambda}\right)^{\alpha}
\left[\log\frac{m}{\Lambda}\right]^{\beta}
\left(\frac{p_{\bot}}{\Lambda}\right)^{\gamma},
\end{equation}
where $N_{\mathrm{int}}$ is the total number of particles in the final and initial states that participate in the 
interaction.  Also $\alpha$, $\beta$, 
and $\gamma$ are non-negative integers.  
For simplicity we display one component of transverse momentum, $p_{\bot}$; however, the general form includes a 
product of all transverse components from all particles.  In principle, the introduction of a particle mass allows 
any function of $\frac{m}{\Lambda}$ to appear.  However, to ${\cal O}(g_{_{\Lambda}}^{3})$ the only extra scale 
dependence comes in the form $\left(\frac{m}{\Lambda}\right)^{\alpha}\left[\log\frac{m}{\Lambda}\right]^{\beta}$.  
If $\beta=0$ and
\begin{equation}
\label{condition}
6-2N_{\mathrm{int}}-\alpha-\gamma=0,
\end{equation}
the term is independent of the cutoff and is referred to as a ``cutoff-independent'' contribution.  These 
contributions are discussed in the next subsection.  

The expression for a matrix element of $V^{(r)}_{\mathrm{CD}}\hspace{-.07cm}(m,\Lambda)$ is derived from 
Eq. (\ref{diff}):
\begin{equation}
\label{fcd}
\left<F \right| V^{(r)}_{\mathrm{CD}}(m,\Lambda) \left| I \right> = \left[ \langle F\vert
\delta V^{(r)}\hspace{-.07cm}(m,\Lambda,\Lambda^{\prime})\vert I\rangle
 - \sum_{s=2}^{r-1}  B_{r-s,s}
\left<F \vert V^{(r-s)}(m,\Lambda) \vert I \right>\right]_{\Lambda \hspace{.1cm}\mathrm{terms}} .
\end{equation}
``$\Lambda$ terms'' means the terms in the momentum and mass expansion that contain $\Lambda^{\prime}$ are to be removed 
from the expression in brackets.  In terms that depend on 
positive powers of $\Lambda^{\prime}$, we do this by letting $\Lambda^{\prime}\rightarrow 0$, and in terms that 
depend on negative powers of $\Lambda^{\prime}$, we let $\Lambda^{\prime}\rightarrow \infty$.

\subsection{Cutoff-Independent Contributions to $\bbox{V^{(r)}\hspace{-.07cm}(m,\Lambda)}$}
\label{independent}

Considering the condition in Eq. (\ref{condition}), 
only two-point and three-point interactions can have cutoff-independent contributions.  The lowest-order cutoff-independent 
three-point interaction is $V_{\mathrm{CI}}^{(3)}\hspace{-.07cm}(m)$ and has not been explicitly computed in the 
massless or massive theories.  However,  $V_{\mathrm{CI}}^{(2)}\hspace{-.07cm}(m)$ is the lowest-order 
cutoff-independent two-point interaction and must be calculated before anything is calculated to third order. 

The matrix elements of $V_{\mathrm{CI}}^{(r)}\hspace{-.07cm}(m)$ are divided into 2-point and 3-point 
contributions, and are given by the expression
\begin{eqnarray}
\label{cime}
\langle F\vert V^{(r)}_{\mathrm{CI}}\hspace{-.07cm}(m)\vert I\rangle=\frac{1}{B_{r,2}}\left[\langle F\vert\delta 
V^{(r+2)}\hspace{-.07cm}(m,\Lambda,\Lambda^{\prime})
\vert I\rangle 
-\sum_{s=3}^{r+1}B_{r+2-s,s}\langle F\vert V^{(r+2-s)}\hspace{-.07cm}(m)\vert 
I\rangle\right]^{3-\mathrm{point}}_{m^{0}\vec{p}_{\bot}^{\hspace{.05cm}0}\hspace{.1cm}\mathrm{term}}\nonumber\\
+\frac{1}{B_{r,2}}\left[\langle F\vert\delta V^{(r+2)}\hspace{-.07cm}(m,\Lambda,\Lambda^{\prime})\vert I\rangle 
-\sum_{s=3}^{r+1}B_{r+2-s,s}\langle F\vert V^{(r+2-s)}\hspace{-.07cm}(m)\vert I\rangle
\right]^{\mathrm{2-point}}_{m^{2}\hspace{.1cm}\mathrm{term}}.
\end{eqnarray}
Here, ``$m^{0}\vec{p}_{\bot}^{\hspace{.1cm}0}\hspace{.1cm}\mathrm{term}$'' and ``$m^{2}\hspace{.1cm}\mathrm{term}$''
means expand the term in brackets in powers of external transverse momenta and in powers and logs of $m$, and keep only 
the term that is proportional to $m^{0}\vec{p}_{\bot}^{\hspace{.1cm}0}$ or $m^{2}$, respectively.\footnote{The 
2-point contribution is independent of $\vec{p}_{\bot}$ because of cluster decomposition, transverse rotational 
invariance, and boost invariance.}  The removal of $\Lambda$ and $\Lambda'$ dependence is guaranteed by construction.

Initially Eq. (\ref{cime}) looks useless because $V_{\mathrm{CI}}^{(r)}\hspace{-.07cm}(m)$ depends on 
$V^{(r+1)}_{\mathrm{CI}}\hspace{-.07cm}(m)$ [which is inside an integral in $\delta V^{(r+2)}$], suggesting the theory 
must be solved to all orders simultaneously.  However, contributions to the reduced interaction from three-point 
interactions can only appear at odd orders, and contributions from two-point interactions can appear only at even orders.  
Thus, in the massless theory, this apparent problem does not manifest itself because there are no cutoff-independent two-point 
interactions.   In the massive theory, although there are cutoff-independent two-point interactions, it is possible to 
solve for $V^{(2)}_{\mathrm{CI}}\hspace{-.07cm}(m)$ and $V^{(3)}_{\mathrm{CI}}\hspace{-.07cm}(m)$ simultaneously, without 
considering higher orders.  This even-order/odd-order solution pattern can be extended to all orders.  

Including self-energy contributions, the theory we want to describe contains particles of mass $m$.  We can simplify 
the problem by using this fact instead of using Eq. (\ref{cime}) to solve for the even-order 
$V_{\mathrm{CI}}^{(r)}\hspace{-.07cm}(m)$'s.  We do this by forcing the completely disconnected 
parts of the forward $T$-matrix elements to be zero.  
(This part of a $T$-matrix element contains initial and final states that have the same number of particles, 
$n$, and $n$ momentum-conserving delta functions.)  This fixes the even-order 
$V_{\mathrm{CI}}^{(r)}\hspace{-.07cm}(m)$'s since they only involve interactions on single particle lines.  
This allows us to calculate 
$V^{(2)}_{\mathrm{CI}}\hspace{-.07cm}(m)$ independently of $V^{(3)}_{\mathrm{CI}}\hspace{-.07cm}(m)$.  This extra 
condition can be used to fix all even-order $V_{\mathrm{CI}}^{(r)}\hspace{-.07cm}(m)$'s.

\section{Results}
\label{results}
The coupling in this theory runs at third order.  We can compare the coupling at two different 
scales, $\Lambda$ and $\Lambda^{\prime}$:
\begin{equation}
g_{_{\Lambda}}=g_{_{\Lambda^{\prime}}}+\sum_{s=3}^{\infty}g_{_{\Lambda^{\prime}}}^{s}\hspace{.2cm}
\hspace{-.12cm}C_{s}(m,\Lambda,\Lambda^{\prime}).
\end{equation}
We can determine how the coupling runs at third order by solving for $C_{3}(m,\Lambda,\Lambda^{\prime})$ 
(which is proportional to the matrix element 
$\langle\phi_{2}\phi_{3}\vert\delta V^{(3)}\hspace{-.07cm}(m,\Lambda,\Lambda^{\prime})\vert\phi_{1}\rangle\vert_{p_{2}=
p_{3}}$).  Figure \ref{fig:c3} shows how $C_{3}(m,\Lambda,\Lambda^{\prime})$ depends on the mass. The running of the 
coupling is exponentially damped as the mass grows since the cutoff inhibits production of 
intermediate particles.  The difference between the values of the running coupling at two different scales increases 
as the two scales are separated.  This is shown by the 
larger magnitude of $C_{3}(m,\Lambda,\Lambda^{\prime})$ as the separation between $\Lambda$ and $\Lambda^{\prime}$ 
grows. 

\begin{figure}[h]
\unitlength1in
\begin{minipage}[t]{3.1in}
\begin{picture}(3.1,3)
\put(-0.05,.1){\epsfig{file=fig1.epsf,width=3.0in}}
\put(2.0,2.45){{\footnotesize $\Lambda^{\prime}=2.2$}}
\put(2.0,2.30){{\footnotesize $\Lambda^{\prime}=3.0$}}
\put(1.5,-.05){{\footnotesize $m/\Lambda$}}
\put(0,1.2){\rotatebox{90}{{\footnotesize $C_{3}(m,\Lambda,\Lambda^{\prime})$}}}
\put(2.04,2.0){{\footnotesize $\Lambda=2.0$}}
\end{picture}
\par

\caption{The third-order coefficient of the running coupling as a function of the particle mass.  Curves for various 
upper cutoffs with fixed lower cutoff show the coupling is exponentially damped with increasing mass.}
\label{fig:c3}
\end{minipage}
\hfill
\begin{minipage}[t]{3.15in}
\begin{picture}(3.1,3)
\put(0,.1){\epsfig{file=fig2.epsf,width=3.0in}}
\put(-.15,1.1){ \rotatebox{90}{{\small $\frac{\langle \phi_{2} \phi_{3}\vert {\cal 
 M}^{2}_{\mathrm{NC}}(g_{_{\Lambda}},m,\Lambda)\vert\phi_{1}\rangle}{\delta_{1,23}
 \hspace{.1cm}g_{_{\Lambda}}}$}}}
\put(2.05,.76){{\footnotesize $m/\Lambda=0.005$}}
\put(2.05,.58){{\footnotesize $m/\Lambda=0.335$}}
\put(2.05,.4){{\footnotesize $m/\Lambda=0.545$}}
\put(2.25,2.3){{\footnotesize $y=\frac{1}{2}$}}
\put(2.17,2.5){{\footnotesize $g_{_{\Lambda}}^{2}=\frac{512 \pi^{3}}{3}$}}
\put(1.65,-.05){{\footnotesize $\vert\vec{k}_{\bot}\vert/\Lambda$}}
\put(2.23,2.1){{\footnotesize $\Lambda=2.0$}}
\end{picture}
\par
\caption{The matrix element of the non-canonical part of the invariant-mass operator for 
$\phi_{1}\rightarrow\phi_{2}\phi_{3}$ versus the magnitude of the transverse momentum 
in the center-of-momentum frame.  $y$ is the longitudinal momentum fraction carried by particle 2.}
 \label{v3}
\end{minipage}
\end{figure}
Determining $V_{\mathrm{CI}}^{(3)}\hspace{-.07cm}(m)$ requires a fifth-order calculation and is not attempted.  However, 
calculating the matrix element $\langle\phi_{2}\phi_{3}\vert V_{\mathrm{CD}}^{(3)}\hspace{-.07cm}(m,\Lambda)\vert\phi_{1}\rangle$ 
gives the relative sizes of the non-canonical interactions and the canonical interaction.  Their relative magnitudes 
are similar to those in Ref. \cite{brent}, suggesting that an expansion of the reduced interaction in powers of the 
running coupling is valid through third order.

Figure \ref{v3} shows how the non-canonical part of the matrix element of the invariant-mass operator for the 
interaction $\phi_{1}\rightarrow\phi_{2}\phi_{3}$ depends on the magnitude of the transverse momentum in the 
center-of-momentum frame.  Increasing the transverse momentum in the center-of-momentum frame increases the free mass of the 
system.

This work was partially supported by National Science Foundation grant PHY-9800964.

\end{document}